\documentclass{PoS}

\usepackage{psfrag}
\usepackage{graphicx,amsfonts,amsmath,amsthm,mathrsfs,amsopn,bm,bbold}

\usepackage{epsfig}
\def\text{{}}   \newcommand{\tr}{{\rm Tr}}
\def\lsim{\raise0.3ex\hbox{$<$\kern-0.75em\raise-1.1ex\hbox{$\sim$}}}
\def\gsim{\raise0.3ex\hbox{$>$\kern-0.75em\raise-1.1ex\hbox{$\sim$}}}

\newcommand{\lla}{\left\langle}
\newcommand{\rla}{\right\rangle}

\title{Renormalization of Polyakov loops in fundamental and higher representations}

\ShortTitle{Renormalization of Polyakov loops}

\author{\speaker{Olaf Kaczmarek}\\
   Fakult\"{a}t f\"{u}r Physik,
 Universit\"{a}t Bielefeld, D-33615 Bielefeld, Germany\\
        E-mail: \email{okacz@physik.uni-bielefeld.de}
}

\author{Sourendu Gupta\\
Department of Theoretical Physics, Tata Institute for Fundamental Research,\\
 Homi Bhabha Road, Mumbai 400 005, India}

\author{Kay H\"{u}bner\\
Physics Department, Brookhaven Natl. Laboratory, Upton, New York 11973, USA}

\date{\today}

\abstract{
We compare two renormalization procedures, one based on the short
distance behavior of heavy quark-antiquark free energies and the other by using
bare Polyakov loops at different temporal extent of the lattice and find that
both prescriptions are equivalent, resulting in renormalization constants that
depend on the bare coupling. Furthermore these renormalization constants show
Casimir scaling for higher representations of the Polyakov loops.\\
The analysis of Polyakov loops in different representations of
the color SU(3) group indicates that a simple perturbative inspired
relation in terms of the quadratic Casimir operator is realized to a good
approximation at temperatures $T \gsim T_c$ for renormalized as well as bare
loops.\\
In contrast to a vanishing Polyakov loop in representations with non-zero
triality in the confined phase, the adjoint loops are small but non-zero even
for temperatures below the critical one. The adjoint quark-antiquark pairs
exhibit screening. This behavior can be related to the binding energy of
gluelump states.
}

\FullConference{The XXV International Symposium on Lattice Field Theory\\
		 July 30-4 August 2007\\
		 Regensburg, Germany}

\begin{document}
\section{Introduction}
Studies of the transition from a confined to a deconfined medium as well
as the fundamental question for a proof of confinement are strongly related to
the Polyakov loop. Models based on the Polyakov loop are proposed to describe
the transition to a quark gluon plasma phase and its properties at zero as well
as non-zero baryon density in a phenomenological manner
\cite{Pisarski:2006hz,Dumitru:2003hp,Dumitru:2004gd,
    Dumitru:2005ng,Dumitru:2003cf,Megias:2005ve,Megias:2004hj,Ratti:2005jh,
    Meisinger:2003id,Fukushima:2003fm,Diakonov:2004kc}.
Furthermore the connection of SU(3) theory to the large $N_c$-limit (in a
mean-field approximation) is widely discussed
\cite{Dumitru:2003hp,Dumitru:2004gd}.\\
For a test of the reliability and comparison of these models to pure gauge
theory and QCD with dynamical quarks, a detailed knowledge of the behavior of
the renormalized Polyakov loop in the fundamental and higher representations in
those theories is of fundamental importance.\\
We will present two different renormalization procedures for the Polyakov loop
for different representations, show their equivalence and discuss our main
results of this study in pure SU(3) gauge theory.
\section{Fundamental and adjoint Polyakov loops}
\begin{figure}[t]
\epsfig{file=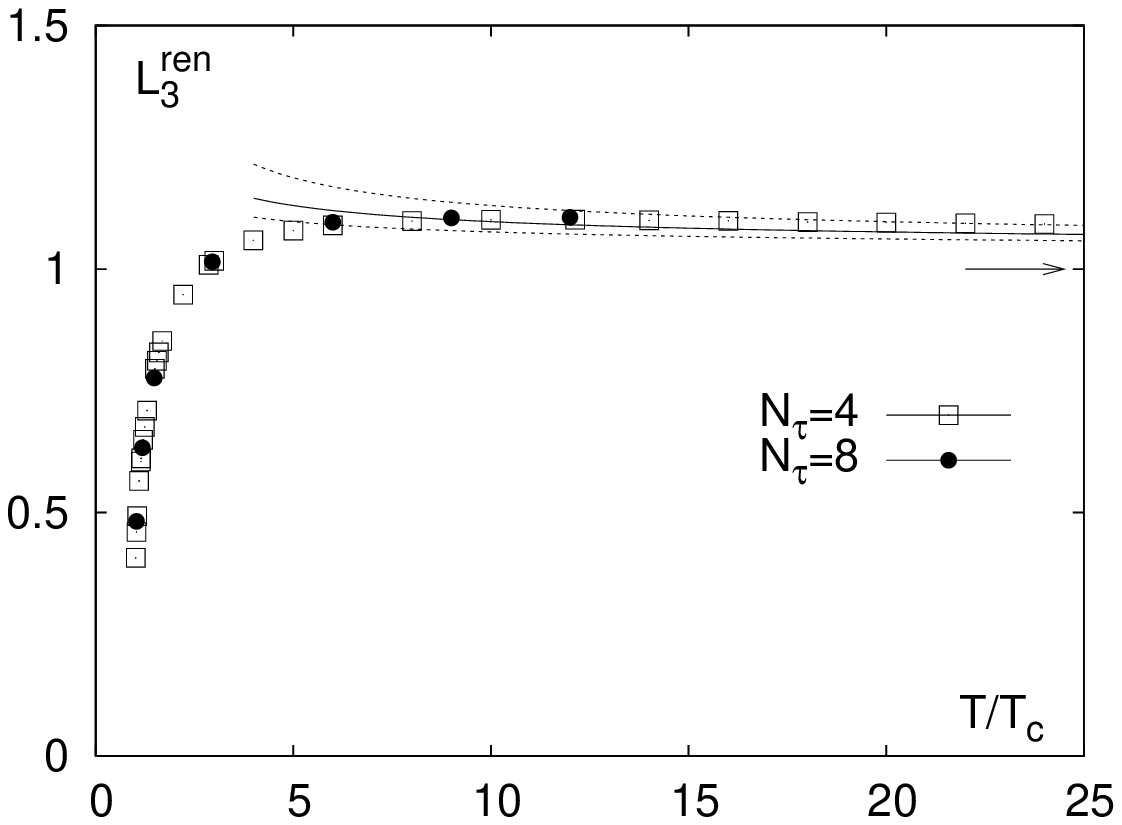,width=7.5cm}
\epsfig{file=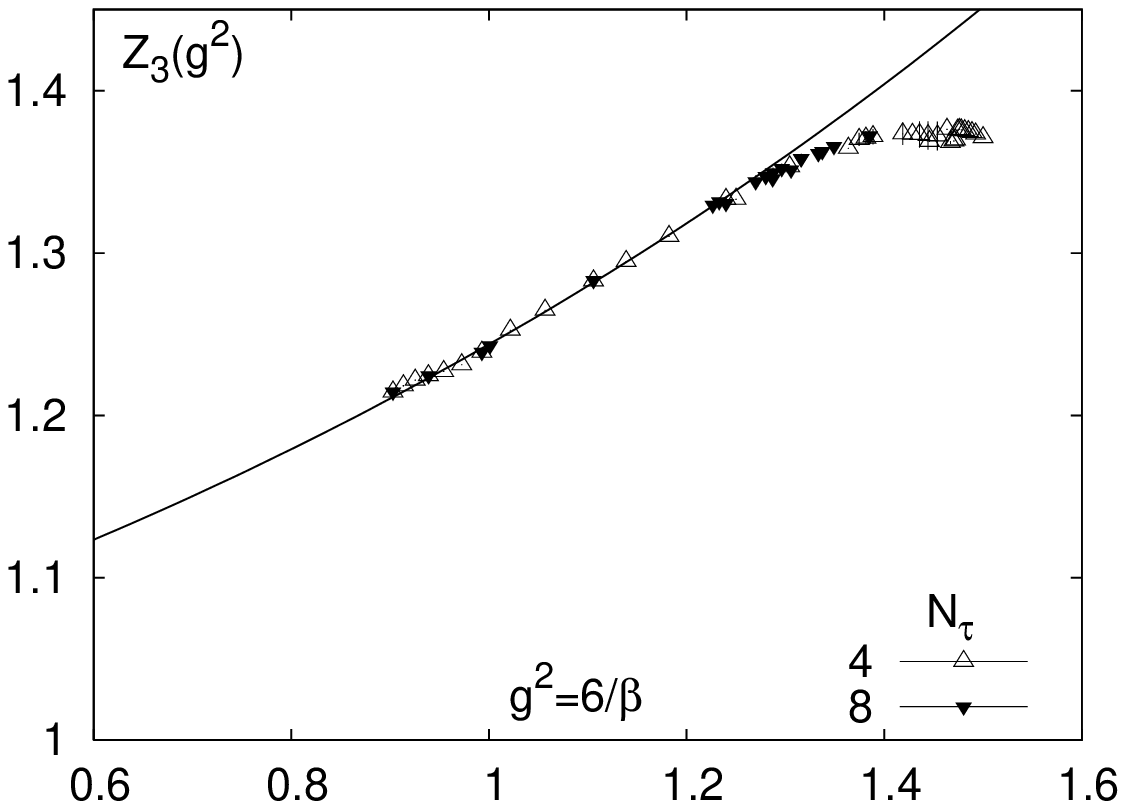,width=7.5cm}
\caption{Renormalized fundamental Polyakov loop (left) and renormalization
  constants (right) in SU(3) pure gauge theory for
  two values of the temporal lattice extent $N_\tau$. The lines in the left
  figure show the
  perturbative result \cite{Gava:1981qd,Polyakov:1980ca} .The arrow represents the
  asymptotic high temperature limit, $L^{ren}=1$. The line in the right figure
  shows a perturbative inspired fit.}
\label{fig1}
\end{figure}
The renormalization of Polyakov loops (in the fundamental representation)
using the short distance behavior of static quark-antiquark free energies was
outlined in \cite{Kaczmarek:2002mc}. For arbitrary representations of the static sources this can
be written as,
\begin{eqnarray}
e^{-F_D^1(r,T)/T} = \left(Z_D(g^2)\right)^{2 d_D N_\tau} \langle \tr (L_D(\vec x)
L_{D}^{\dagger}(\vec y)),
\end{eqnarray}
which is equivalent to the renormalization of the Polyakov loop itself,
\begin{eqnarray}
L^{ren}_D  = \left(Z_D(g^2)\right)^{N_\tau d_D} \langle L^{bare}_
D
\rangle.
\label{lren}
\end{eqnarray}
The renormalization constants are obtained by matching the free energies to the zero temperature
potential at short distances. In fig.~\ref{fig1} we show the results for the renormalized
Polyakov loop (left) and the renormalization constants (right) for two different
$N_\tau$ obtained in quenched QCD. The good agreement of $Z_3(g^2)$ and $L_3(T)$ for the
$N_\tau=4$ and 8 indeed shows that the renormalization constants depend only on
the bare coupling constants.
In perturbation theory Casimir scaling for heavy quark potentials is realized (at
least) up to two-loop order \cite{Schroder:1998vy,Bali:2002wf}. 
\section{Direct renormalization in higher representations}
\begin{figure}[t]
  \psfrag{g}[c]{$\scriptstyle g$ \scriptsize fixed}
  \psfrag{n1}[c]{$\scriptstyle N_{\tau,1}$}
  \psfrag{n2}[c]{$\scriptstyle N_{\tau,2}$}
  \psfrag{t1}[c]{$\scriptstyle T_{\text{start}}$}
  \psfrag{t2}[c]{$\scriptstyle T_{1}$}
  \psfrag{lr}[l]{$\scriptstyle L_D^R\,(T)$}
  \psfrag{l}[l]{$\scriptstyle \lla | L_D|\rla(g^2,N_\tau)$}
  \hspace*{0.4cm}\epsfig{file=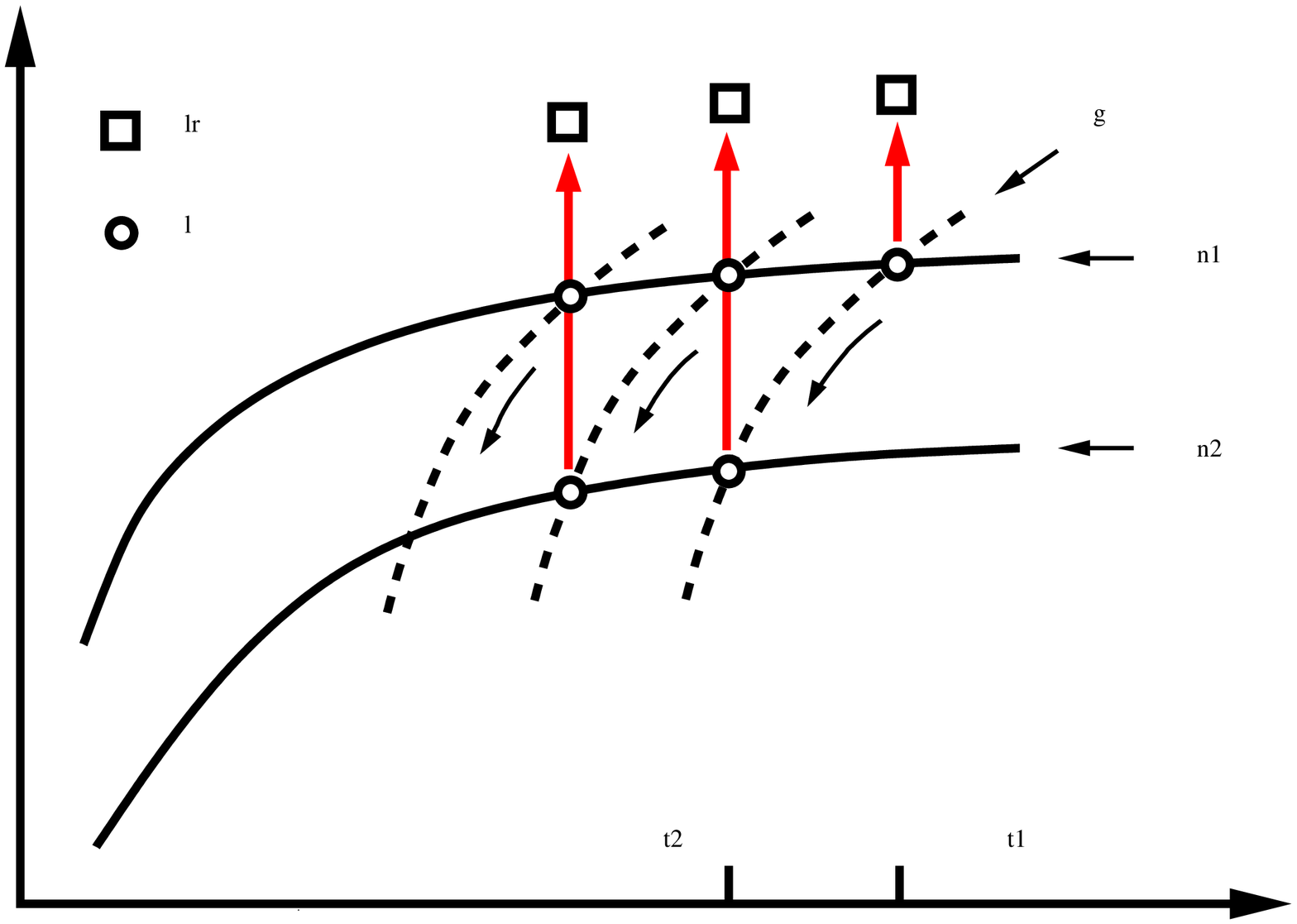,width=6.5cm}
  \hspace*{0.4cm}\epsfig{file=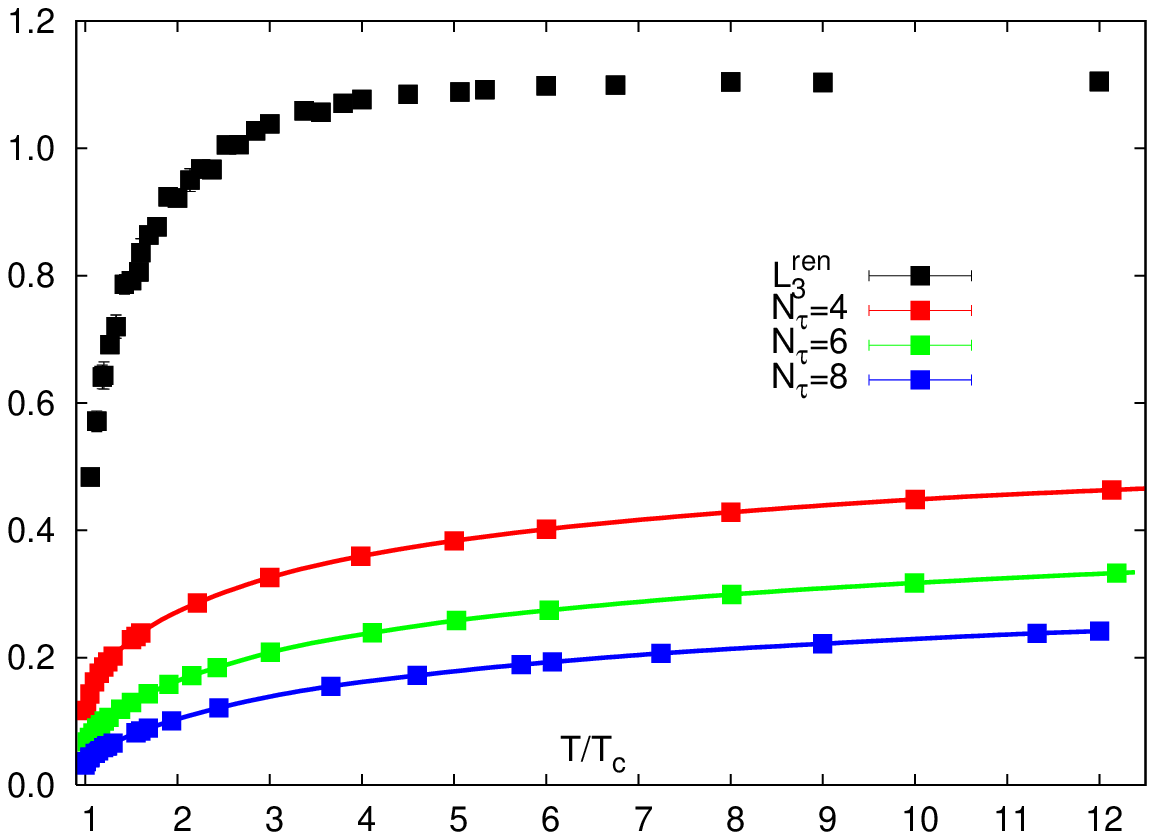,width=7.5cm}
\caption{Renormalization procedure using different $N_\tau$ (left). Bare
  Polyakov loops from $32^3\times N_\tau$ lattices and the resulting
  $L_3^{ren}$. The lines are spline interpolations (right).}
\label{fig2}
\end{figure}
\begin{figure}[t]
\epsfig{file=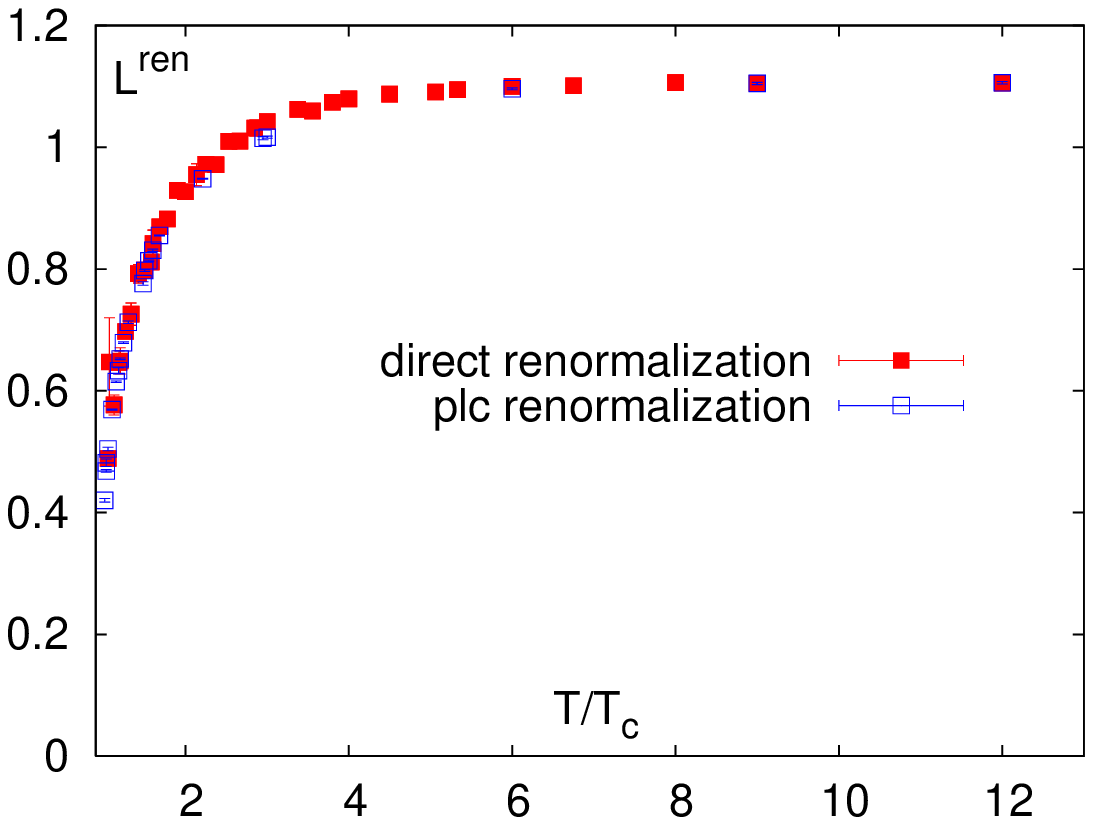,width=7.5cm}
\epsfig{file=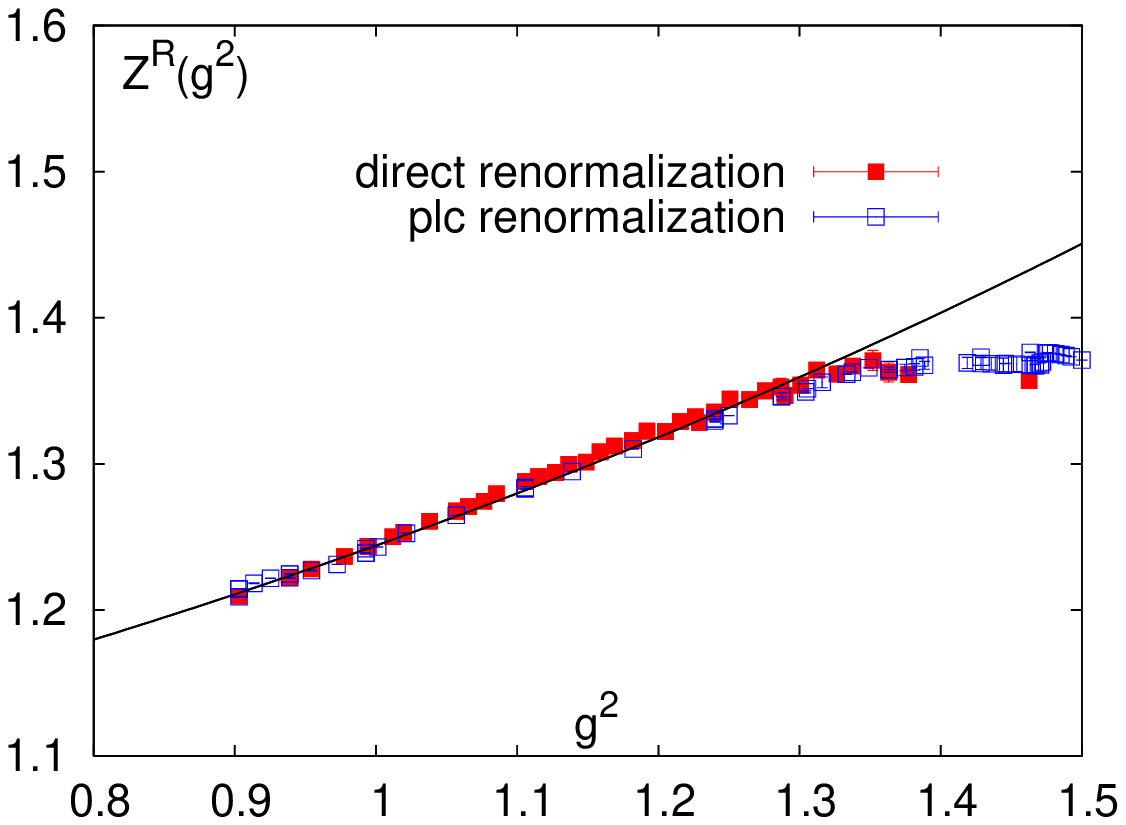,width=7.5cm}
\caption{Comparison of the renormalized Polyakov loop (left) and the
  renormalization constants (right) obtained with the two different
  renormalization procedures.}
\label{fig3}
\end{figure}

\begin{figure}[t]
\epsfig{file=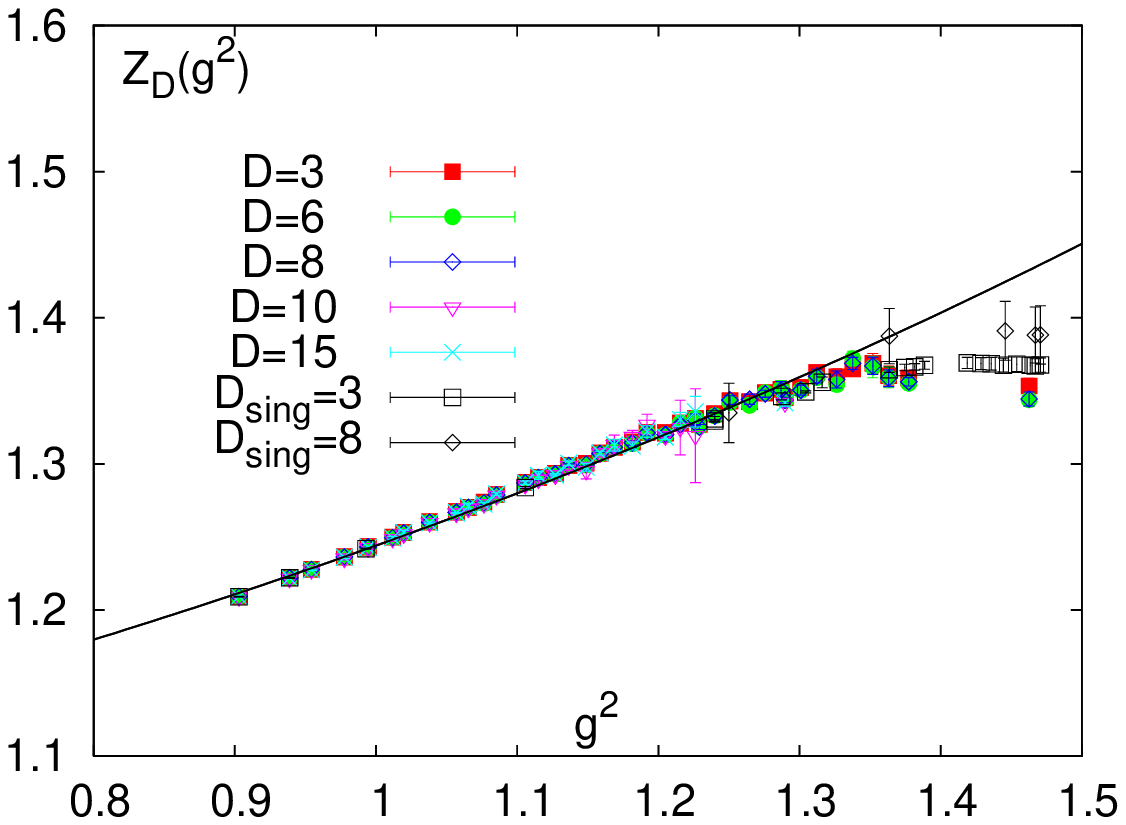,width=7.5cm}
\epsfig{file=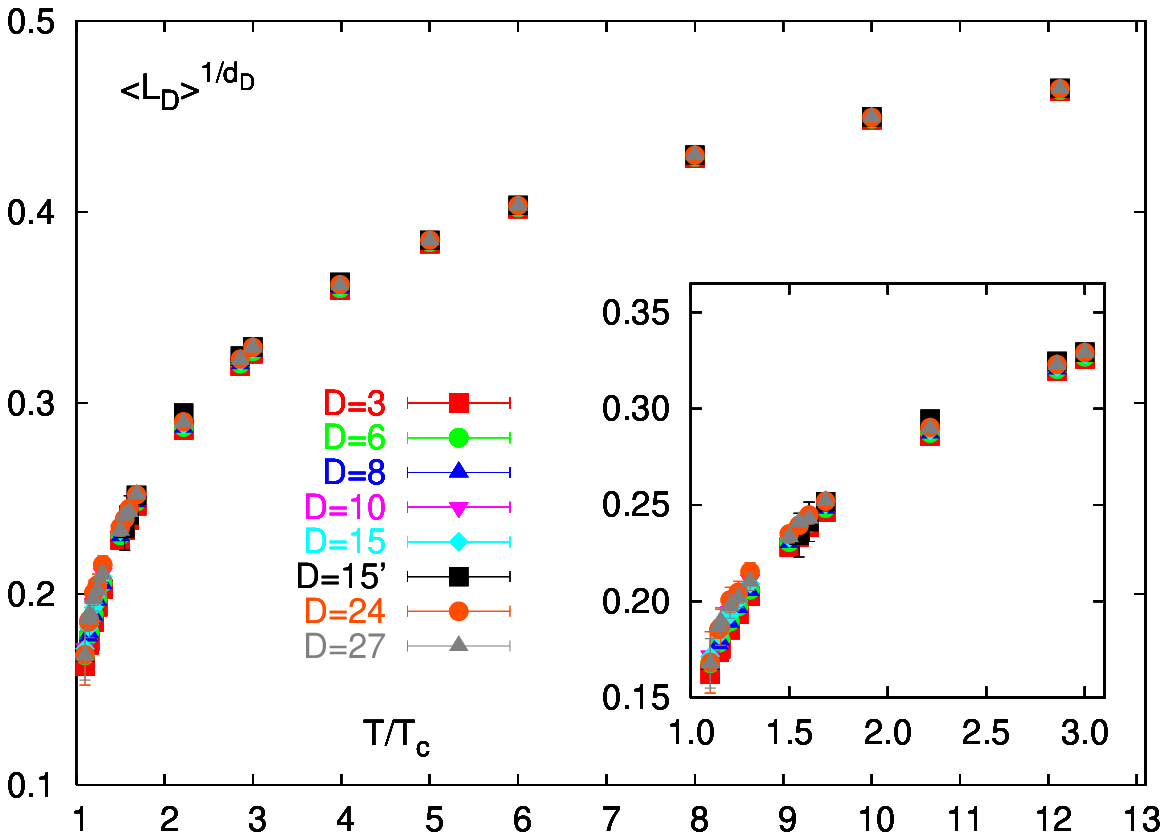,width=7.5cm}
\caption{
Renormalization constants obtained with the direct renormalization procedure
(left).
Also shown are the results obtained from
the previous method for fundamental and adjoint loops, labeled $D_{sing}$.
Casimir scaled bare Polyakov loops (right) for
different representations $D$.
}
\label{fig:renconst}
\end{figure}

Using the observation that the renormalization constants depend only on the
bare couplings opens the possibility for a direct renormalization procedure
based on single bare Polyakov loops at different $N_\tau$ rather than using
Polyakov loop correlation functions
(a similar method was proposed in \cite{Creutz:1980hb}).\\
The fist step in this procedure is to fix the arbitrary overall scale factor by
fixing the value of the renormalized Polyakov loop at the highest temperature
in our analysis, $T_i=T_{max}=12 T_c$, where we use the same scheme as in the
previous method. From this we obtain the renormalization constants at the
corresponding coupling (at two different $N_\tau$) by assuming
\begin{eqnarray}
Z_D^{d_DN_{\tau,i}}(g^2_{i})
L^{bare}_D(g^2_{i},N_{\tau,i})\vert_{\frac{1}{a(g^2_{i})N_{\tau,i}} = T_i}
&\equiv& L^{ren}_D(T_i) \ \ \ 
\mathrm{and} \\ 
Z_D^{d_DN_{\tau,j}}(g^2_{j})
L^{bare}_D(g^2_{j},N_{\tau,j})\vert_{\frac{1}{a(g^2_{j})N_{\tau,j}} = T_i}
&\equiv& L^{ren}_D(T_i).
\end{eqnarray}
This procedure can now be iterated (see fig.~\ref{fig2}~(left)) to obtain the
renormalization constants and the renormalized Polyakov loop down to $T_c$.
In fig.~\ref{fig2}(right) we show the result of this procedure for the
fundamental loop in SU(3) pure gauge theory obtained by applying this procedure
for three values of $N_\tau$.\\
The comparison of the two renormalization procedures (fig.~\ref{fig3}) indeed
shows that the
renormalized Polyakov loops (left) and the renormalization constants (right)
are in good
agreement and both procedures are equivalent.\\
The prescription can easily be extended to Polyakov loops in any
representation $D$ \cite{Gupta:2006qm,Gupta08}, thus giving the renormalized Polyakov loops
$L^R_D$
and the renormalization constants $Z_D(g^2)$. 
Using these,
one can then check Casimir scaling in the form 
\begin{eqnarray}
Z_D(g^2) =
Z_3(g^2),
\label{Zcas}
\end{eqnarray}
for the renormalization constants and
\begin{eqnarray}
L_D^{ren}(T) = \left(L_3^{ren}(T)\right)^{d_D}
\label{lcas}
\end{eqnarray}
for the Polyakov loops,
where $d_D=C_2(D)/C_2(3)$ is the
ratio of quadratic Casimirs. 
The test of Casimir scaling is then the independence of $Z$ from $D$.
Note that (\ref{Zcas}) together with (\ref{lren}) implies that, if Casimir scaling
is realized for the renormalized Polyakov loop, it holds for the bare loops as
well.\\
In fig.~\ref{fig:renconst} we show the results for the renormalization
constants (left) and the Casimir scaled bare Polyakov loops (right) for
representations up to
$D=15$. For comparison we also include the results obtained from the
previous method. We observe a good agreement of $Z_D$ for all
representations in the whole coupling range and for the scaled Polyakov loops
for temperatures down to the critical one. 
\section{Adjoint Polyakov loops and string breaking}
\begin{figure}[t]
            \epsfig{file=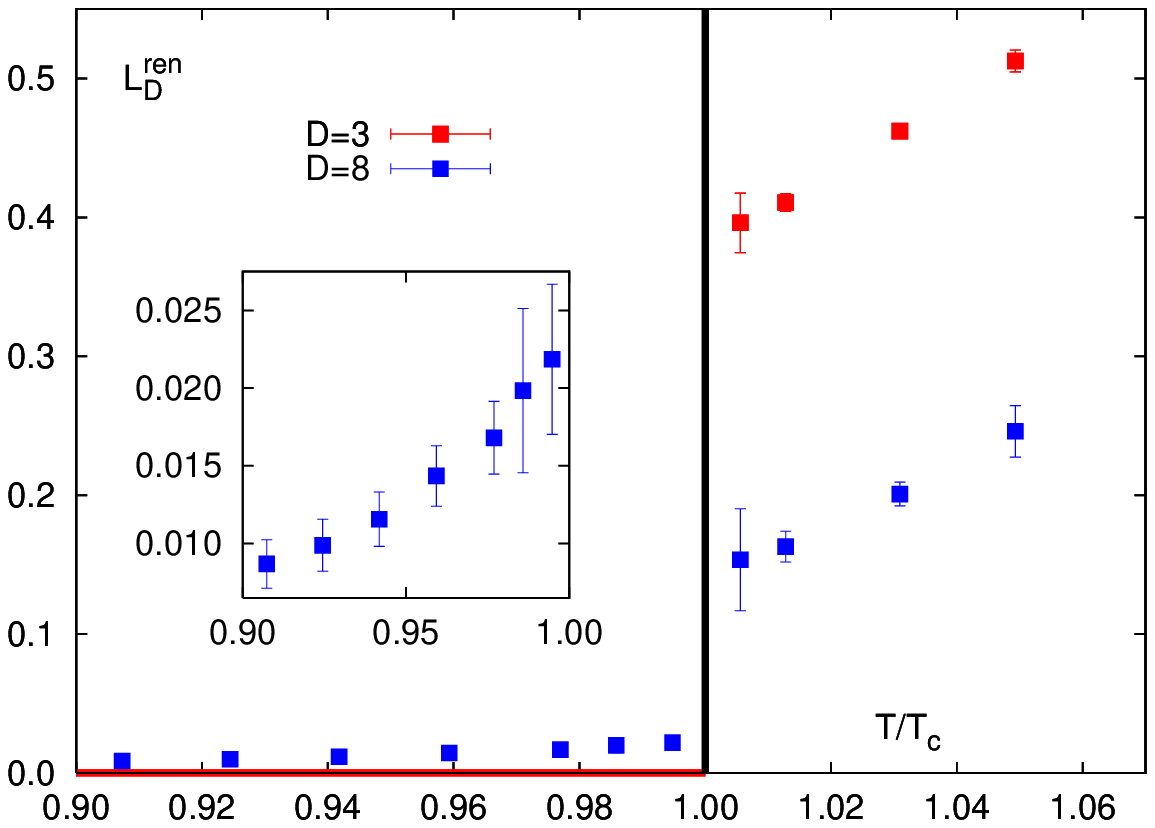, width=7.5cm}
            \epsfig{file=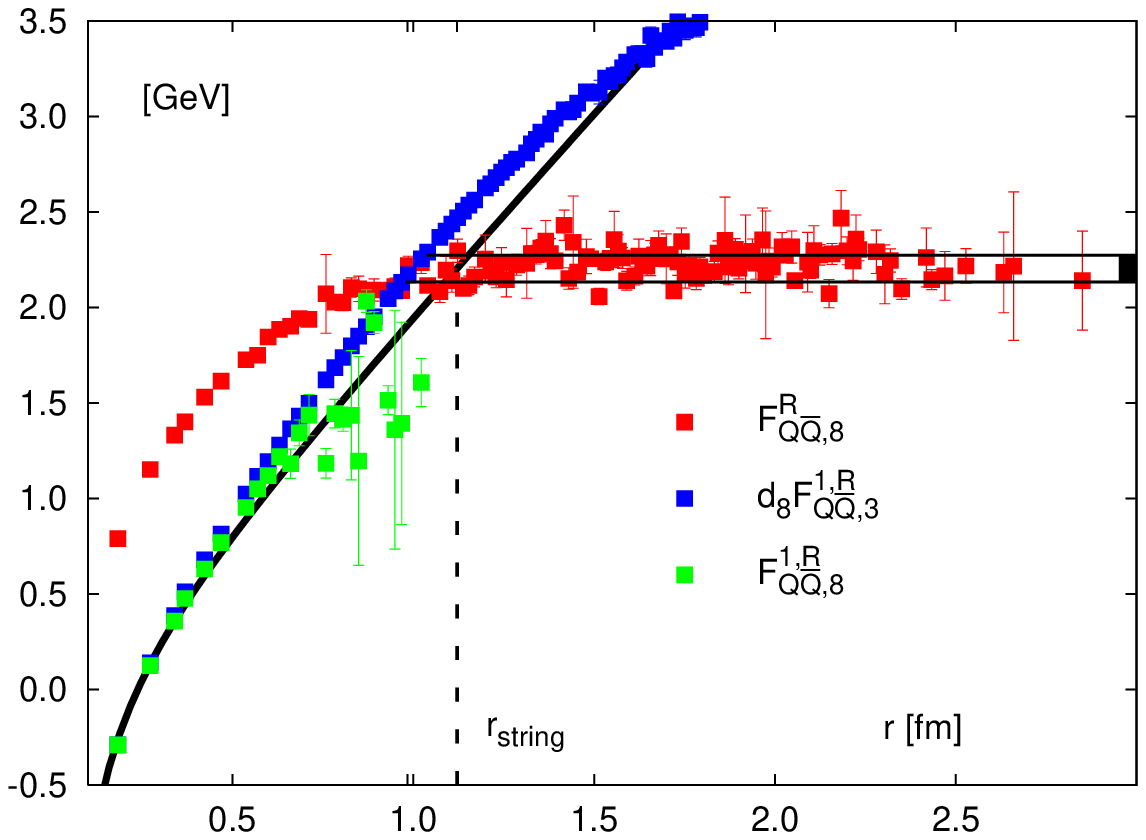, width=7.5cm}
\caption{Renormalized adjoint Polyakov loop compared to the fundamental
  loops (left). Heavy quark-antiquark free energies for adjoint sources in the
  color singlet and color averaged channel compared to Casimir scaled color
  singlet free energy of fundamental sources at $T=0.959~T_c$. The lines show
  the asymptotic value and estimates for the string breaking distance.}
\label{fig4}
\end{figure}

\begin{figure}[t]
            \center\epsfig{file=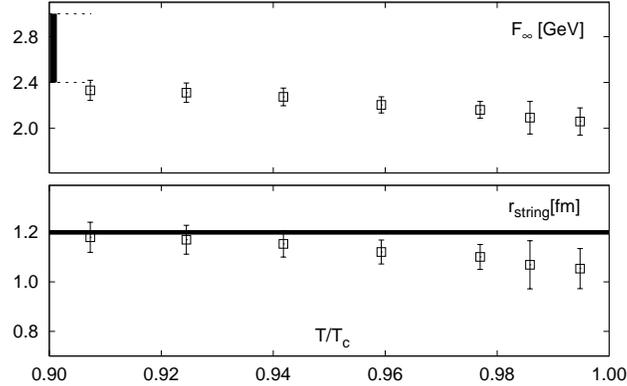, width=8.7cm}
\caption{Asymptotic value of the adjoint heavy quark free energies (upper
  panel). Estimate of the string breaking radius (lower panel).}
\label{fig5}
\end{figure}

In contrast to Polyakov loops with non-zero triality, which have
vanishing expectation values in the confined phase (in the infinite
volume limit), 
for all triality-zero representations (r=8,10,27,...) one expects to see
string breaking below $T_c$ also in pure gauge theory, 
and hence a non-vanishing Polyakov loop in the infinite
volume limit
(see also discussions in 
\cite{Damgaard:1987wh,Redlich:1988ig,Fingberg:1990ag}).\\
We have computed the infinite volume, renormalized adjoint Polyakov
loop below $T_c$. 
Fig.~\ref{fig4}~(left) shows the results compared to the fundamental loop
around $T_c$. While the fundamental renormalized Polyakov
loop is zero below $T_c$, the adjoint loop is small but clearly
non-vanishing.\\
For the other triality-zero representations ($r=10,27$) we expect the
same behavior, but we cannot
give the infinite volume limit below $T_c$, since the corresponding
data is still too noisy for the
statistics achieved in this work.\\
For the heavy quark-antiquark free energies of adjoint sources we
observe string breaking below $T_c$ (fig.~\ref{fig4}~(right)). 
The asymptotic value of the static quark-antiquark free energy of
adjoint sources (fig.~\ref{fig5}) can be related to the binding energy of
gluelump
states \cite{Bali:2003jq}, i.e. bound states of a dynamical gluon with a static adjoint
source. In the upper part of fig.~\ref{fig5} we show the results for the
asymptotic values of the adjoint heavy quark free energies and in the lower
part an estimate for the string-breaking radius defined through
\begin{eqnarray}
V_8(r_{string}) = F_8(r=\infty,T),
\end{eqnarray}
where $V_8$ is the zero temperature potential and $F_\infty$ is the asymptotic value
of the quark-antiquark free energy, both for adjoint sources.\\ 
An extension of this study will be the analysis of color octet states
of heavy quark-antiquark free energies with fundamental sources
combined with a static adjoint source, forming a color singlet state
in total. 
\section{Conclusions and Outlook}
We have extended the renormalization procedure outlined in \cite{Kaczmarek:2002mc} 
to quark sources in the adjoint
representation. We observed that the resulting renormalization constants only
depend on the bare coupling $g^2$. This led to the proposal of a new (direct)
renormalization procedure for the Polyakov loop itself measured at different
temporal lattice extent in contrast to the (indirect) renormalization using
two-point correlation functions of Wilson lines or Polyakov loops.\\
We have shown that both procedures are indeed equivalent leading to a solid
description of the renormalized Polyakov loop. Furthermore we applied the new
prescription to
Polyakov loops in the fundamental and higher representations up to $D=27$.\\
The direct renormalization procedure is solely based on gauge invariant
quantities, while the $q\bar q$-renormalization is based on color singlet
correlation functions of Wilson lines which are (in principle) gauge dependent
quantities. The equivalence of both procedures, i.e. the agreement of the
renormalization constants and the renormalized Polyakov loops, shows that (at
least) the short (temperature independent) as well as the (asymptotic) large
distance part of the heavy quark free energies obtained in Coulomb gauge become
gauge independent as proposed in \cite{Philipsen:2002az,Jahn:2004qr}.\\
The analysis of Polyakov loops in higher representations up
to $D=27$ led to the
the observation that Casimir scaling for the Polyakov loops and the
corresponding renormalization constants in different representations is a
surprisingly good approximation even down close to $T_c$. This may indicate that
non-Casimir scaling
terms in a perturbative series may only play a sub-dominant role.\\
Due to the Z(3)-symmetry of the pure gauge theory, all Polyakov loops with
non-zero triality vanish in the confined phase even in the absence of dynamical
quarks, i.e. pure gauge theory. 
For the adjoint representation
we have observed small, but non-zero values below $T_c$. The static adjoint
sources can couple to the dynamical adjoint constituents (gluons) of the theory
and the quark-antiquark pair gets screened even in the confined phase. This
screening phenomenon ({\it string breaking}) is visible in the heavy quark free
energies which have a finite asymptotic value while for zero-zero triality they
rise linearly with distance. The finite asymptotic value for adjoint sources
may be related to the binding energy of gluelump
states.\\
A more detailed study and discussion of the renormalization of Polyakov loops
in higher representations as well as the application to QCD with dynamical
quarks is in preparation \cite{Gupta08}.
A future extension of this study will be the analysis of correlation functions
of different representations, e.g. a baryonic system made up of a color octet
state of a quark-antiquark pair in the fundamental representation combined with a
static adjoint source forming a color singlet state in total.
\section*{Acknowledgment}
This work has been supported by 
Contract No. DE-AC02-98CH10886 with the U.~S.~Department of Energy.
At an early stage of this work
K.~H.~ has been supported by the DFG under grant GRK 881/1.
S.~G. would like to acknowledge the hospitality of the University of
Bielefeld.

\end{document}